\documentclass[twocolumn]{aastex631}
\usepackage{graphicx}
\usepackage{float}
\usepackage{amsmath}
\usepackage{amssymb}
\usepackage{subfigure}
\usepackage{cases}
\usepackage{mathrsfs}
\usepackage[flushleft]{threeparttable}
\usepackage{amssymb}
\usepackage{pifont}
\usepackage{epstopdf}
\usepackage{xcolor}
\usepackage[all]{hypcap}
\usepackage{mwe,tikz}
\usepackage{bm}
\usepackage{tabularx}
\usepackage{multirow}
\usepackage{longtable}
\usepackage{xspace}
\usepackage{rotating}
\usepackage{CJK}
\usepackage{upgreek}
\usepackage{booktabs}
\usepackage{textcomp}

\newcommand{\e}{\text{e}}

\newcommand{\p}{\partial}

\begin{document}

 \title{GRB 211211A-like Events and How Gravitational Waves May Tell Their Origins}
\correspondingauthor{Bin-Bin Zhang}
\email{bbzhang@nju.edu.cn}

\author[0000-0002-5596-5059]{Yi-Han Iris Yin}
\affiliation{School of Astronomy and Space Science, Nanjing University, Nanjing 210093, China}
\affiliation{School of Physics, Nanjing University, Nanjing 210093, China}

\author[0000-0003-4111-5958]{Bin-Bin Zhang}
\affiliation{School of Astronomy and Space Science, Nanjing University, Nanjing 210093, China}
\affiliation{Key Laboratory of Modern Astronomy and Astrophysics (Nanjing University), Ministry of Education, China}
\affiliation{Purple Mountain Observatory, Chinese Academy of Sciences, Nanjing, 210023, China}

\author[0000-0002-9615-1481]{Hui Sun}
\affiliation{{Key Laboratory of Space Astronomy and Technology, National Astronomical Observatories, Chinese Academy of Sciences, Beijing 100012, China}}

\author[0000-0002-5485-5042]{Jun Yang}
\affiliation{School of Astronomy and Space Science, Nanjing University, Nanjing 210093, China}
\affiliation{Key Laboratory of Modern Astronomy and Astrophysics (Nanjing University), Ministry of Education, China}

\author[0000-0001-7402-4927]{Yacheng Kang}
\affiliation{Department of Astronomy, School of Physics, Peking University, Beijing 100871, China}
\affiliation{Kavli Institute for Astronomy and Astrophysics, Peking University, Beijing
100871, China}

\author[0000-0002-1334-8853]{Lijing Shao}
\affiliation{Kavli Institute for Astronomy and Astrophysics, Peking University, Beijing
100871, China}
\affiliation{National Astronomical Observatories, Chinese Academy of Sciences, Beijing
100012, China}

\author[0000-0003-0691-6688]{Yu-Han Yang}
\affiliation{Department of Physics, University of Rome ``Tor Vergata'', via della Ricerca Scientifica 1, I-00133 Rome, Italy}

\author[0000-0002-9725-2524]{Bing Zhang}
\affiliation{Nevada Center for Astrophysics, University of Nevada Las Vegas, NV 89154, USA}
\affiliation{Department of Physics and Astronomy, University of Nevada Las Vegas, NV 89154, USA}

\begin{abstract} 
 
GRB 211211A is a rare burst with a genuinely long duration, yet its prominent kilonova association provides compelling evidence that this peculiar burst was the result of a compact binary merger. However, the exact nature of the merging objects, whether they were neutron star pairs, neutron star--black hole systems, or neutron star--white dwarf systems, remains unsettled. This {\it Letter} delves into the rarity of this event and the possibility of using current and next-generation gravitational wave detectors to distinguish between the various types of binary systems. 
Our research reveals an event rate density of $\gtrsim 5.67^{+13.04}_{-4.69} \times 10^{-3}\ \rm Gpc^{-3}\ yr^{-1}$ for GRB 211211A-like gamma-ray bursts (GRBs), which, assuming GRB 211211A is the only example of such a burst, is significantly smaller than that of typical long- and short-GRB populations. We further calculated that if the origin of GRB 211211A is a result of a neutron star--black hole merger, it would be detectable with a significant signal-to-noise ratio (S/N), given the LIGO-Virgo-KAGRA designed sensitivity. On the other hand, a neutron star--white dwarf binary would also produce a considerable S/N during the inspiral phase at decihertz and is detectable by next-generation spaceborne detectors DECIGO and the Big Bang Observer. However, to detect this type of system with millihertz spaceborne detectors like LISA, Taiji, and TianQin, the event must be very close, approximately 3 Mpc in distance or smaller.

\end{abstract}

\section{Introduction}
\label{sec:introduction}

Short gamma-ray bursts (GRBs) have long been proposed to originate from the merger events of the binary neutron star system \citep{1989Natur.340..126E,1992ApJ...395L..83N,1998A&A...338..535R,10.1046/j.1365-8711.2002.05898.x} or the neutron star--black hole system \citep{1991AcA....41..257P,1998ApJ...494L..53K,Janka_1999}, both being prominent gravitational wave emission sources. The codetection of electromagnetic \citep{Abbott_2017, Savchenko_2017, Goldstein_2017} and GW signals \citep[LIGO-Virgo Collaboration;][]{PhysRevLett.119.161101} during the GW 170817/GRB 170817A event has provided compelling evidence linking a short GRB to its merger origin. However, establishing a comprehensive sample that can determine whether all short GRBs stem from mergers and whether all binary neutron star (BNS)/neutron star--black hole (NS-BH) GW events are associated with short GRBs is still far from being achieved. In addition, mounting evidence, including the identification of the core-collapse type short-duration GRB 200826A \citep{ZhangBB2021NA} and the compact-star-merger-type long-duration GRB 211211A \citep{YangJun_2022,Zhang_2022_lat}, implies that a GRB's physical type is more intricate than what can be inferred solely from burst duration \citep{2006Natur.444.1010Z,zhangb2009typeItypeIIpaper}. Thus, studying the potential GW emission from such unusual events, particularly those of merger type, can enhance our comprehension of GW-related phenomena and provide insights into future GW detections.

The peculiar long-duration burst, GRB 211211A, consists of a main emission (ME) phase lasting for $\sim 13$ s and an extended emission (EE) phase lasting for $\sim 55$ s but was proposed to originate from a compact binary merger due to its compelling evidence of kilonova association \citep{YangJun_2022, 2022Natur.612..228T, 2022Natur.612..223R, 2022Natur.612..236M, 2023NatAs...7...67G, Zhu_2022, Gao_2022,Chang_2023, kunert2023model}, although \citet{collapsar} pointed out that a collapsar origin for GRB 211211A could be barely possible. While most studies have been focused on the compact star merger scenarios, the exact composition of the compact binary system remains unresolved at present. \citet{Zhu_2022} showed that the merger between an NS and a BH can roughly reproduce the multiwavelength observations. Meanwhile, \citet{kunert2023model} conducted a Bayesian model-selection study and concluded that a BNS merger scenario is the most likely. However, none of the BNS or NS-BH merger papers have provided a satisfactory interpretation to the long duration of the burst. \citet{YangJun_2022} pointed out that a neutron star--white dwarf (NS-WD) merger involving a massive WD component near the Chandrasekhar limit and a similarly massive NS component is self-consistent with all the observations, ranging from prompt gamma-rays and early X-ray afterglow to the engine-fed kilonova emission. This scenario is supported by a more detailed modeling by \citet{zhong2023grb}. We noticed that the scenario involving an NS-WD merger was not included in the analysis by \citet{kunert2023model}, which only compared the BNS merger, NS-BH merger, core-collapse supernova, and collapsar scenarios.

One key feature that can differentiate between the NS-WD merger scenario and the BNS or NS-BH merger scenarios is through their associated GW signals, as they dominate different frequency ranges. Specifically, the NS-BH merger produces a GW signal at around $\sim$100 Hz, whereas the NS-WD merger typically produces signals at around 0.1 Hz. In this \textit{Letter}, we investigate whether future ground-based and spaceborne GW detectors, operating across the wave band from millihertz to kilohertz, can differentiate between the two distinct types of GW signals and thereby determine the nature of events similar to GRB 211211A.

The paper is structured as follows. In \S \ref{sec:rate}, we first present the event rate density and further calculated how rare the burst is assuming the GRB originates from NS-WD. In \S \ref{sec:signal}, we calculate the predicted GW signals in the frequency domain for NS-BH and NS-WD mergers, respectively. In \S \ref{sec:detection}, after introducing future GW missions and detectors that are representative of the current state-of-the-art and future developments in the field, we calculate the detecting capabilities and derive the matched filtering signal-to-noise ratio (S/N) of two different types of GW signals during their inspiral and merger phases. Finally, we conclude with a brief summary and discussion in \S \ref{sec:sum}.

\section{How Rare Is GRB 211211A?}
\label{sec:rate}

Prior research, including studies by \cite{YangJun_2022, 2022Natur.612..228T, 2022Natur.612..223R, 2022Natur.612..236M, 2023NatAs...7...67G, Zhu_2022, Gao_2022}, has indicated that this event is highly uncommon among long GRBs. GRB 211211A is a nearby bright GRB at $z\ =\ 0.0763$ \citep{2022Natur.612..223R} but could be detectable up to a maximum redshift $z_{\rm max}=0.52$ (see below). The detection of GRB 211211A suggests that the number of such events within the volume enclosed by $z_{\rm max}$ is at least 1 but could be more. We assume that the detected number of such type of burst by Fermi Gamma-ray Burst Monitor (GBM) is $N\ \gtrsim\ 1$. Following \citet{2015ApJ...812...33S}, one can estimate the local event rate density $ \rho_{\rm 0,GRB 211211A}$ through
\begin{equation}
\label{equ:erd}
\frac{\Omega T}{4\pi}\rho_{\rm 0,GRB 211211A}V_{\rm max}= N \gtrsim 1 ,
\end{equation}
where $\Omega = 8\ {\rm sr}$ is the solid angle for the field of view of GBM. The effective operational time for the period between 2008 and approximately 2022 is denoted by $T = 7\ {\rm yr}$. This is calculated based on the working time of GBM since 2008, assuming a duty cycle of approximately 50\%.
\begin{equation}
\label{equ:vmax}
V_{\rm max} = \int_{0}^{z_{\rm max}}\frac{f(z)}{1+z}\frac{dV(z)}{dz}dz
\end{equation}
is the maximum comoving volume, and $V(z)$ is redshift-dependent specific comoving volume based on {standard cosmological parameters, i.e., $\rm H_{0} = 69.6\ {\rm km}\ s^{-1}\ Mpc^{-1}$, $\Omega_{M}=0.286$, $\Omega_{vac}=0.714$ \citep{2014ApJ...794..135B}, so}
\begin{equation}
\label{equ:dvdz}
\frac{dV(z)}{dz}=\frac{c}{H_0}\frac{4\pi D_{L}^{2}}{(1+z)^{2}[\Omega_{M}(1+z)^{3}+\Omega_{vac}]^{1/2}}.
\end{equation}

The dimensionless redshift distribution function $f(z)$ describes the redshift-dependent merger event rate density, i.e.,
\begin{equation}
\label{equ:local}
\rho(z) = \rho_{0}f(z).
\end{equation}
Here, we adopt the simulated NS-WD event rate density {$\rho(z)$} from a consistent stellar population synthesis paradigm \citep{2019MNRAS.482..870E} and derive an empirical expression of $f(z)$ {from Eq. \eqref{equ:local}} using multiple-power law (PL) functions as \citep{2008ApJ...683L...5Y,2015ApJ...812...33S}
\begin{align}
f(z)= 
& \bigg[(1+z)^{1.638\eta}+\Big(\frac{1+z}{6.381}\Big)^{-3.134\eta} \\
& +\Big(\frac{1+z}{11.176}\Big)^{-1.173\eta}\bigg]^{1/\eta}
\end{align}
where $\eta$ is set to be $-3.482$.

\begin{figure}
 \centering
 \includegraphics[width = 0.43\textwidth]{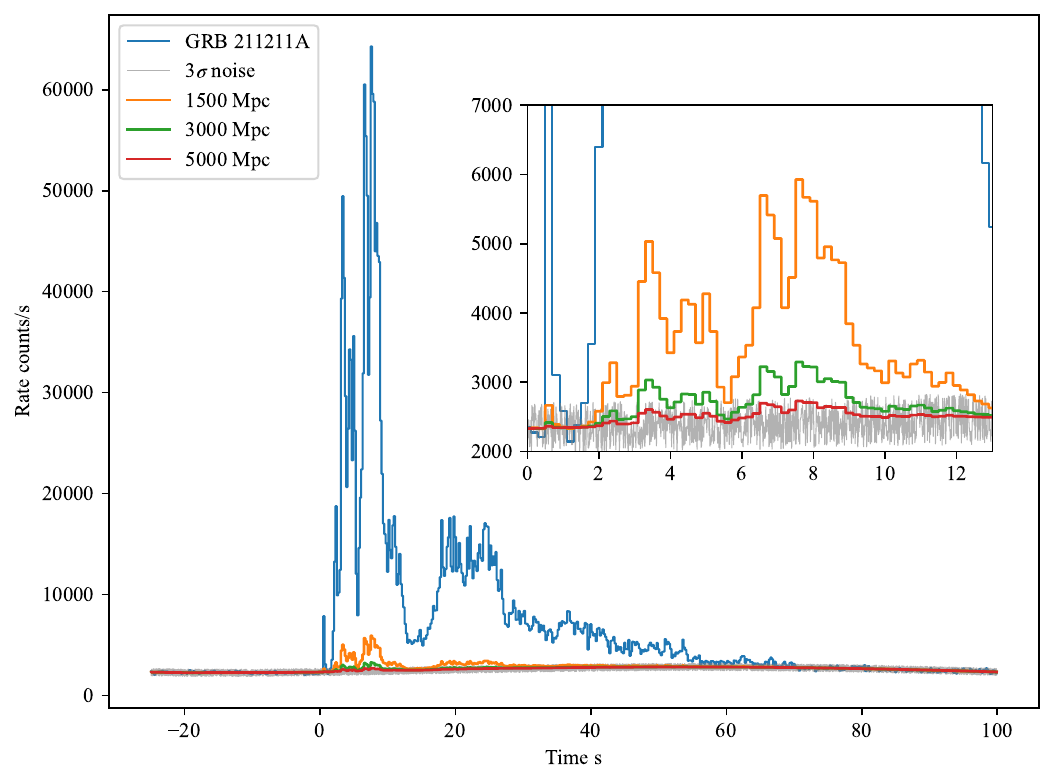}
 \caption{The observed light curve of GRB 211211A and pseudo light curves by putting the burst at different distances. The background light curve (gray) is incorporated with a 3$\sigma$ noise level derived from observation. At a redshift $z \sim 0.52$ (or $3000\ {\rm Mpc}$), the total duration of the pseudo burst is $\sim 6\ {\rm s}$. At a redshift $z \sim 0.79$ (or $5000\ {\rm Mpc}$), the light curve of the pseudo burst becomes indistinguishable from 3$\sigma$ noise \citep[see also][]{2022Natur.612..228T}.}
 \label{fig:distance}
\end{figure}

The maximum redshift $z_{max}$ in Eq. \eqref{equ:vmax} is determined by the farthest distance within which this event can be observed. Additionally, we require that at $z_{max}$, GRB 211211A would still appear as a genuinely long-duration burst to maintain its rarity \citep{YangJun_2022}. Under such two conditions, we calculated $z_{max}$ by placing the burst at different distances and identifying a critical value of $D_{L}=3000\ {\rm Mpc}$ (which corresponds to $z_{\rm max}=0.52$), as illustrated in Figure \ref{fig:distance}. 
{At $D_{L}=3000\ {\rm Mpc}$, GRB 211211A would still be classified as a long GRB with $T_{\rm 90} \sim 6$ s, and the associated supernova detection \citep[e.g.,][]{ 2022Natur.612..223R,2022Natur.612..228T} would remain feasible, the lack of which we could consider as a distinctive characteristic for merger origin.}

\begin{figure}
 \centering
 \includegraphics[width = 0.48\textwidth]{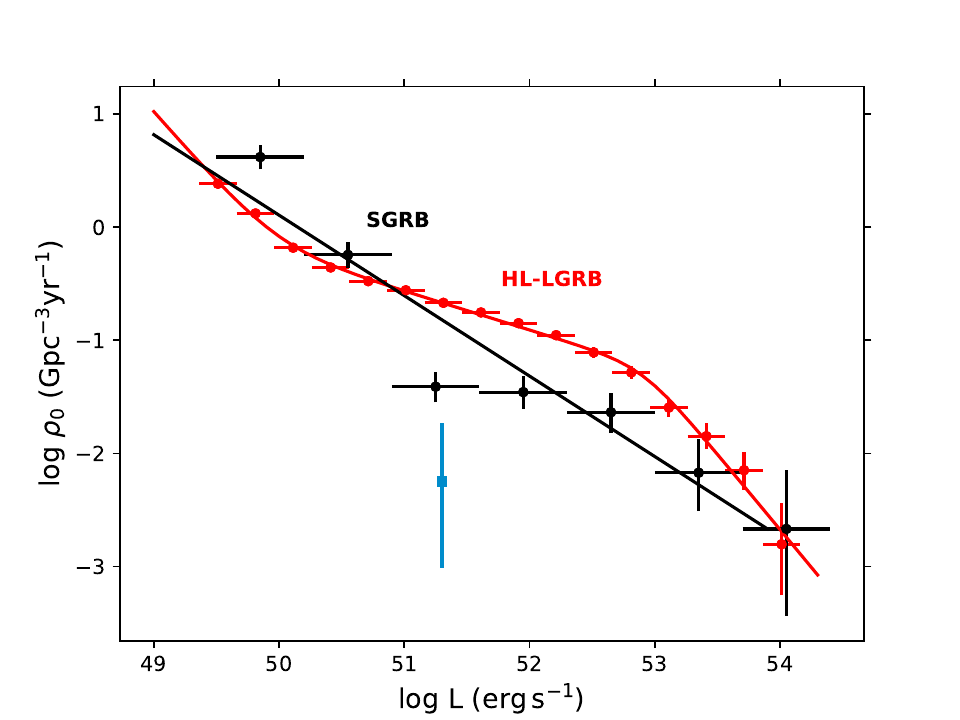}
\caption{Event rate density for short (black) and long (red) GRBs, alongside the value of GRB 211211A (blue). The solid curves in black and red represent the empirical fits obtained from \cite{2015ApJ...812...33S}. The vertical error bar indicates the 1$\sigma$ Gaussian errors obtained from \citet{1986ApJ...303..336G}.}
 \label{fig:rate}
\end{figure}

Substituting $z_{\rm max}=0.52$ to Eq. (\ref{equ:vmax}), we obtain the event rate density of GRB 211211A from Eq. (\ref{equ:erd}) as
\begin{equation}
\label{eq:rate}
\rho_{\rm 0,GRB211211A} = 5.67^{+13.04}_{-4.69} \times 10^{-3}N\ \rm Gpc^{-3} yr^{-1}.
\end{equation}
With a conservative value of $N=1$, we plot the event rate density of Eq. \ref{eq:rate} together with those of the short- and long-GRB populations, as depicted in Figure \ref{fig:rate}. It is crucial to exercise caution when interpreting this inference, as it is subject to significant uncertainty of $N$. A precise determination of $N$ necessitates a comprehensive investigation into potential concealed events within archival GRB data or the future detection of analogous events. These aspects, unfortunately, fall outside the scope of this paper.

Regardless of the high uncertainties of $N$, one can still gauge the rarity of the event by considering the fraction of NS-WD systems capable of producing a GRB 211211A-like event, supposing that it originates from a special type of NS-WD merger. Assuming that the cosmological distributions of BNS and NS-WD systems mirror those in our Galaxy, we calculated a ratio between the two kinds of systems according to the census by \cite{Pol_2021}, namely, there are $\sim 5000$ Galactic ultracompact BNS systems and $\sim 6700$ Galactic ultracompact NS-WD systems. Hence the ratio yields $k=\frac{6700}{5000}=1.34$. We then applied this factor to the cosmological beamed short GRB event rate density adopted from \citet{2015ApJ...812...33S}, taking a typical minimum luminosity $\sim 10^{50}\ \rm erg\ s^{-1}$, acquiring $\rho_{\rm BNS}{\rm\ (beamed)} \sim 1.3\ \rm Gpc^{-3}\ yr^{-1}$ for the Gaussian delay model. Under the assumption of the same beaming factor for NS-WD GRB systems, the resultant calculation yielded a cosmological beamed event rate density of NS-WD systems $\rho_{\rm NS-WD}{\rm\ (beamed)}=k\times \rho_{\rm BNS}{\rm\ (beamed)}=1.74\ \rm Gpc^{-3}\ yr^{-1}$, if all NS-WD mergers make GRBs. We note that such a value aligns with calculations obtained from other NS-WD population synthesis models \citep[e.g.,][]{2009arXiv0912.0009T,2018A&A...619A..53T} when the beaming factor is taken into account within the calculations.
With both $\rho_{\rm 0,GRB211211A}$ and $\rho_{\rm NS-WD}{\rm\ (beamed)}$, one can estimate the fraction of NS-WD mergers that indeed make GRBs, i.e.,
\begin{align}
f =
& \frac{\rho_{\rm 0,GRB211211A}}{\rho_{\rm NS-WD}{\rm\ (beamed)}} = \frac{5.67 \times 10^{-3}N\ \rm Gpc^{-3}\ yr^{-1}}{1.74\ \rm Gpc^{-3}\ yr^{-1}} \nonumber \\
& \approx \frac{N}{307}\simeq 3.26N \text{\textperthousand}.
\end{align}
With an $N\sim$ a few, one can see that $f$ is a very small value, suggesting that only a small fraction of NS-WD systems are capable of producing GRBs \citep{YangJun_2022}. This finding is in agreement with the calculation in \citet{zhong2023grb}. Physically, \cite{YangJun_2022} argued that only WDs near the Chandrasekhar limit could induce an accretion-induced collapse upon merger, making a rapidly spinning magnetar that can power the long-duration GRB, its extended X-ray plateau emission, as well as the kilonova emission.

\section{The Gravitational Wave Signal}
\label{sec:signal}

To further distinguish the origin of the burst, we can utilize the differences in the associated GW signals between NS-WD and NS-BH mergers. In this section, we first calculate such signals of the two kinds of systems.

The construction of the NS-BH merger waveform involves the use of two different models. Firstly, we use the TaylorF2 model \citep{PhysRevD.44.3819,PhysRevD.59.124016,2014LRR....17....2B,Faye_2012,PhysRevD.103.123541} in the frequency domain to calculate the observed strain at 347.8 Mpc {\citep[corresponding to redshift $z=0.0763$;][]{2022Natur.612..223R}} during the inspiral phase in the low-frequency range ($<$10 Hz). Next, for the higher-frequency range, we employ the IMRPhenomNSBH model \citep{PhysRevD.101.124059}, which is specifically designed for spinning, nonprecessing NS-BH binaries, and combines post-Newtonian/BH perturbation analytical results with numerical-relativity simulations to represent the GW signals in the relatively high-frequency domain. In both models, the masses of the NS and BH are set to be approximate $\sim 1.23_{-0.07}^{+0.06} M_{\bigodot}$ and $\sim 8.21_{-0.75}^{+0.77} M_{\bigodot}$, respectively \citep{Zhu_2022}. The coherence of the selected models is demonstrated in Figure \ref{fig:signals}, where the two segments (marked as orange dashed and dotted lines) of the GW signal using these models align seamlessly.

The NS-WD merger waveform was generated by following the same approach as mentioned above for the inspiral phase using the TaylorF2 model. However, we truncated the signal at the point where the two stars made contact, which occurred at a distance equal to the sum of their respective stellar radii. For the NS with a mass of 1.4 $M_{\bigodot}$, we used a standard radius of 10 km, while for the 1.3 $M_{\bigodot}$ WD, we referred to the mass-radius relationship outlined in \citet{PhysRevD.84.084007} to obtain a radius of 3.3 $\times$ $10^3$ km. The NS-WD system's merger frequency in the detector's frame is approximately $\sim$ $4.66\times10^{-1}$ Hz. However, due to a limited understanding of the process, no known templates are available to precisely describe the phase after the inspiral stage. We caution that in our waveform, matter and tidal effects from the WD are ignored. The calculated inspiral waveform for the NS-WD merger is presented in Figure \ref{fig:signals} as a blue dashed line. For the merger phase, we utilize the binary BH merger waveform (with BH masses set to the WD and NS masses) as a rough approximation for the signal amplitude, as demonstrated by the faint blue dashed line in Figure \ref{fig:signals}. However, the realistic waveform for this stage could be much more complicated and is out of the scope of this work. It should be noted that this approximation for the merger phase lacks precision and is not taken into account when calculating the S/N.

\section{The Gravitational Wave Detection}
\label{sec:detection}

We now investigate whether GW detectors, if available, are capable of detecting the GW signals resulting from the merger of NS-BH or NS-WD.

\subsection{Sensitivity Curve of Gravitational Wave Detectors}
\label{subsec:curves}

The sensitivity of GW detectors in the frequency domain is represented through the power spectral density (PSD) curve. An analysis of the PSD is conducted for the two categories listed as follows:

\begin{itemize}
 \item Ground-based GW detectors. There are two Advanced LIGO detectors (H1 and L1) in the USA \citep{Aasi_2015}, an Advanced Virgo detector (V1) in Europe \citep{Acernese_2015}, and a KAGRA detector (K1) in Japan \citep{PhysRevD.88.043007,2019NatAs...3...35K}. It is anticipated that LIGO, Virgo and KAGRA will attain their designed sensitivities \citep{2020LRR....23....3A} after O4, which is set to start in 2023 May and plan to observe for 18 calendar months, subject to intermittent one- or two-month intervals for maintenance. Subsequently, the upgraded detectors, referred to as Advanced LIGO Plus (A+), Advanced Virgo Plus (AdV+), and KAGRA+, will undergo further sensitivity improvement. 

 \item Spaceborne GW detectors. In the research of the early evolution of the inspiral phase preceding the merger, spaceborne detectors exhibit promising capabilities in exploring the lower-frequency regime down to millihertz \citep{gao2021midfrequency}. However, considering the stochastic GWs radiated from undetectable sources persisting at a certain level, the imprint of the sources remains for extended periods for spaceborne detectors. Therefore, two planned decihertz detectors, the Decihertz Interferometer Gravitational wave Observatory (DECIGO) mission from Japan \citep{Kawamura_2006} and the Big Bang Observer (BBO), a follow-on mission to LISA from the United States \citep{Harry_2006}, are extremely beneficial in identifying primordial GW background and achieving large S/Ns, for their higher-frequency part is free from those double WD binary confusion noises whose typical frequencies are below 0.2 Hz \citep{2011PhRvD..83d4011Y,dwdbg_ref}. Besides, in exploring Galactic WD binaries and massive BH binaries signals in the millihertz band, three spaceborne GW missions, LISA \citep{Robson_2019},
 Taiji \citep{LUO2020102918}, and TianQin \citep{2016CQGra..33c5010L}, have gradually taken shape. After the test on the pioneering LISA Pathfinder mission, the three missions are hopeful in achieving the designed target sensitivity.
 
\end{itemize}

\begin{figure*}
 \centering
 \includegraphics[width = 0.90\textwidth]{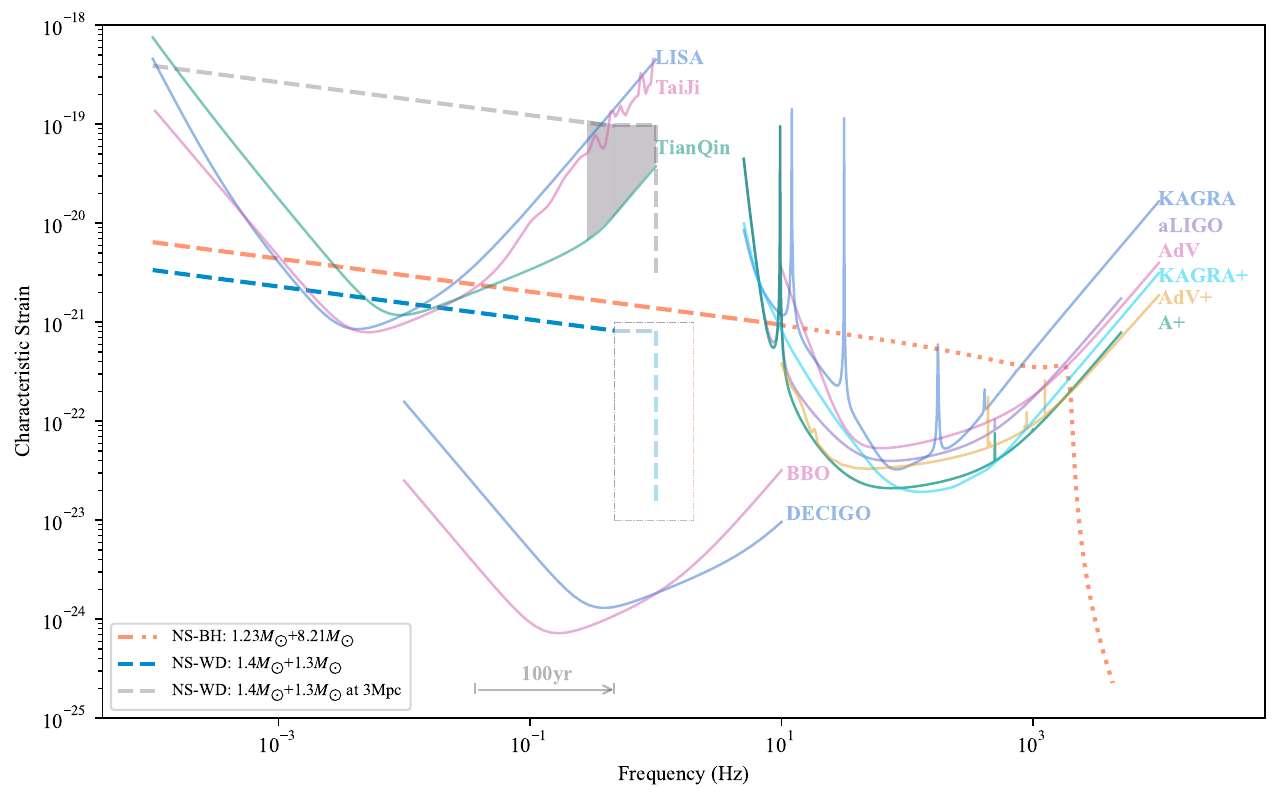}
 \caption{Detection of GRB 211211A GW events using both ground-based and spaceborne detectors. \textit{Orange dashed and dot lines:} the characteristic strain for the case of a NS-BH merger. \textit{Blue dashed lines:} the characteristic strain for the case of a NS-WD merger. \textit{Gray dashed line:} the characteristic strain for the case of a NS-WD merger at 3 Mpc. \textit{Colored solid curves:} the characteristic noise strain of different GW detectors. \textit{Color-filled blocks:} the detecting frequency range for the case of a NS-WD merger at 3 Mpc with TianQin. The gray arrow denotes the duration of the inspiral phase of the NS-WD system, which corresponds to the time taken for the frequency to change from $3.63 \times 10^{-2}$ Hz to $4.66 \times 10^{-1}$ Hz.}
 \label{fig:signals}
\end{figure*}

The PSD curves for all the aforementioned detectors have been obtained from publicly available resources, as outlined in Table \ref{tab:combined} and plotted in Figure \ref{fig:signals} in the form of noise strain defined in Eq. (\ref{equ:hn}).

\subsection{Detecting Capability Calculation}

The S/N (denoted as $\rho$) determines the detectability of a GW event. The optimal S/N can be calculated through matched filtering in the frequency domain using
\begin{equation}
\label{equ:snr}
\rho = 2\sqrt{\int_{f_{\rm low}}^{f_{\rm high}}\frac{|\tilde{h}(f)|^2}{S_{n}(f)}df},
\end{equation}
where $S_{n}(f)$ is the one-sided power spectral density of the interferometer noise and $\tilde{h}(f)$ is the waveform in the frequency domain. 

In addition, we can define dimensionless characteristic signal strain $h_c$ and noise strain $h_n$ as

\begin{equation}
\label{equ:hc}
h_{c}(f)=2f|\tilde{h}(f)|,
\end{equation}
and 
\begin{equation}
\label{equ:hn}
h_{n}(f)=\sqrt{fS_{n}(f)}.
\end{equation}

Using these definitions, Eq. \eqref{equ:snr} can be rewritten as
\begin{equation}
\label{equ:final_snr}
\rho = \sqrt{\int \bigg[\frac{h_{c}(f)}{h_{n}(f)}\bigg]^{2}d(\log f)}.
\end{equation}

Eq. \eqref{equ:final_snr} indicates that the area between the source and detector curves reflects the S/N and, therefore, the detectability, as demonstrated in Figure \ref{fig:signals}.

\subsection{Result}

For the NS-BH merger, it is straightforward to calculate the S/N using Eq. \eqref{equ:final_snr} as the inspiral-merger-ringdown GW signal falls into the detecting ranges of ground-based detectors. To derive the S/N for each ground-based detector, we employed the sensitivity curves obtained in \S \ref{subsec:curves} and utilized a conservative frequency range of 20-512 Hz. The results are presented in Table \ref{tab:combined}.

For the NS-WD system, we only consider the signal detectability in the inspiral phase with the spaceborne detectors, which is only well modeled if the NS and WD can be regarded as well-separated bodies that gradually spiral toward one another. The waveform of the NS-WD system is not well-modeled in the merger phase, for which we do not consider when computing S/N.
In calculating the signal S/Ns in the inspiral phase, the detector's longevity limits the frequency variation as the rate of frequency change is relatively insignificant at low frequency \citep{Maggiore:2007ulw}. In our specific case of the NS-WD system, we have calculated that it takes approximately 100 yr for the signal frequency to evolve from $3.63 \times 10^{-2}$ Hz to the merger frequency of around $4.66\times 10^{-1}$ Hz (as shown in Figure \ref{fig:signals}). 
 We simulated various frequency ranges within 100 yr before merger and calculated the corresponding S/Ns, taking into account a rough observation or operating time frame for each detector. Table \ref{tab:combined} presents the frequency range that exhibits the highest S/N for each spaceborne detector, while ensuring this range is reachable within 100 yr prior to the merger.

Based on our calculations in Table \ref{tab:combined}, if the GRB 211211A originated from an NS-BH merger, all upgraded ground-based detectors can detect the significant high-frequency GW signals with S/Ns exceeding a threshold of 8 \citep{2020LRR....23....3A}. 
Our results of NS-BH detection also align with the predictions reported in \citet{2023MNRAS.518.5483S,2022Natur.612..223R}. However, if the GRB originated from an NS-WD merger, only two spaceborne detectors, namely, BBO and DECIGO with very large S/Ns, can detect the associated GW signal. For the other three spaceborne detectors (i.e., LISA, Taiji, and TianQin) to detect the signal, GRB 211211A would have to be much closer, e.g., at a distance of $\sim$ 3 Mpc or closer, as illustrated in Figure \ref{fig:signals} in color-filled blocks. 

\begin{table*}
\centering
\caption{The S/Ns for NS-BH and NS-WD mergers with various detectors}
\label{tab:combined}
\begin{tabular}{c|cccccc}
\hline
\hline
Merger System & Detectors & Longevity & Distance (Mpc) & Range (Hz) & $\rho$ & PSD References \\
\hline
\multirow{6}*{NS-BH} & aLIGO & - & 347.8 & 20-512 & 22.62 & LIGO-T1800044-v5\footnote{https://dcc.ligo.org/LIGO-T1800044/public} \\
&AdV & - & 347.8 & 20-512 & 16.15 & LIGO-P1200087-v48\footnote{https://dcc.ligo.org/LIGO-P1200087}\\
 &KAGRA & - & 347.8 & 20-512 & 19.54 & JGW-T1707038-v9\footnote{https://gwdoc.icrr.u-tokyo.ac.jp/cgi-bin/DocDB/ShowDocument?docid=7038} \\
&A+ & - & 347.8 & 20-512 & 43.35 & LIGO-T1800042-v5\footnote{https://dcc.ligo.org/LIGO-T1800042/public} \\
&AdV+ & - & 347.8 & 20-512 & 29.52 & LIGO-P1200087-v48\footnote{https://dcc.ligo.org/LIGO-P1200087} \\
 &KAGRA+ & - & 347.8 & 20-512 & 38.27 & JGW-T1809537-v6\footnote{https://gwdoc.icrr.u-tokyo.ac.jp/cgi-bin/DocDB/ShowDocument?docid=9537} \\
\hline
\multirow{8}*{NS-WD}&{BBO}&
{$\sim$ 5 yr}&
{347.8}&
{$(1.03-4.66)\times10^{-1}$}&
{1432.36}&
\multirow{2}*{\citet{2011PhRvD..83d4011Y}}\\
&{DECIGO}&
{$\sim$ 5 yr}&
{347.8}&
{$(1.03-4.66)\times10^{-1}$}&
{635.62}&\\
&\multirow{2}*{LISA}&
\multirow{2}*{$\sim$ 4 yr}&
{347.8}& 
\multirow{2}*{$(3.63-3.70)\times10^{-2}$}& {$4.55\times10^{-2}$}&
\multirow{2}*{\citet{Robson_2019}} \\
&&&{1.98}&&{8}&\\
&\multirow{2}*{Taiji}&
\multirow{2}*{$\sim$ 5 yr}&
{347.8}&
\multirow{2}*{
$(4.58-4.79)\times10^{-2}$}&
{$8.92\times10^{-2}$}&
\multirow{2}*{\citet{taiji}} \\
&&&{3.83}&&{8}&\\
&\multirow{2}*{TianQin}&
\multirow{2}*{\shortstack{2 $\times$ (3 months) \\ each year}}&
{347.8}&
\multirow{2}*{
$(2.84-4.66)\times10^{-1}$}&
{$7.18\times10^{-2}$}&
\multirow{2}*{\citet{2018CQGra..35i5008H}} \\
&&&{3.10}&&{8}&\\
\hline
\hline
\end{tabular}
\end{table*}

\section{Summary and Discussions}
\label{sec:sum}

This {\it Letter} first calculated the event rate density of GRB 211211A and found that only a small portion of the NS-WD system can produce such a burst. This event rate density can be regarded as a conservative lower limit for a long GRB with an NS-WD origin, suggesting that similar events may already be present in the archival data. 

We also examine the possibility of using current and future GW detectors to distinguish the exact compositions of GRB 211211A's origin between the NS-BH system and the NS-WD system. We simulate the predicted GW signals from two approximate compact stars evolution models and calculate the S/Ns of two merger systems signals with future ground-based and spaceborne GW detectors PSD curves. Our results indicate that with LIGO's designed sensitivity, the NS-BH merger that caused GRB 211211A would be detectable with a significant S/N. On the other hand, the NS-WD binary would also generate a notable S/N during the inspiral phase with decihertz spaceborne detectors, such as DECIGO and BBO, but detecting such a system with millihertz spaceborne detectors like LISA, Taiji, and TianQin would require the event to be closer, at approximately 3 Mpc distance.
Given the low event rate derived from integrating Eq. \eqref{eq:rate} over a relatively small volume with $D_{\rm L}\sim 3$ Mpc, it appears more reasonable to prioritize the utilization of next-generation decihertz detectors, such as BBO and DECIGO, for detecting these mergers. Nevertheless, it is important to highlight the potential of millihertz detectors in providing valuable constraints on GRB 211211A-like events, especially when there is nondetection of BNS or NS-BH merger signals in kilohertz. Furthermore, the rate of NS-WD mergers surpasses that of BNS mergers, offering promising prospects for generating early warning alerts for GRB detection \citep{10.1093/mnras/stac1738}, compensating for the low detection rate of GRB 211211A-like events. It is also worth mentioning that our results do not consider the impact of confusion noise nor the matter and tidal effects in NS-WD mergers. Furthermore, our calculation employs an angle-averaged approach and does not consider the location or inclination angles of the GW events. 

Despite the fact that there is a lack of coincident GW and long-GRB detection reported in \citet{Wang_2022}, where they searched for signals from BNS or NS-BH mergers and long GRBs from 4-OGC and the Fermi-GBM/Swift-BAT catalog, our study highlights the potential for further investigation of coincident GW and long-GRB signals from NS-WD mergers with upcoming GW detectors.

\section*{acknowledgements}

We are grateful to Yi-Fan Wang for the valuable discussions. We acknowledge the support by the National Key Research and Development Programs of China (2018YFA0404204, 2022YFF0711404, 2022SKA0130102), the National SKA Program of China (2022SKA0130100), the National Natural Science Foundation of China (grant Nos. 11833003, U2038105, U1831135, 12121003), the science research grants from the China Manned Space Project with NO.CMS-CSST-2021-B11, the Fundamental Research Funds for the Central Universities and the Program for Innovative Talents and Entrepreneur in Jiangsu.

\end{document}